\documentclass[twocolumn,showpacs,preprintnumbers,amsmath,amssymb]{revtex4-1}

\usepackage{graphicx}
\usepackage{dcolumn}
\usepackage{bm}

\begin{document}


\title{Electronic, magnetic and transport properties of
 Fe intercalated 2H-TaS$_2$ studied by means of the KKR-CPA method}
\author{S.\ Mankovsky}
\email{Sergiy.Mankovskyy@cup.uni-muenchen.de}
\affiliation{%
  Department  Chemie,  Physikalische  Chemie,  Universit\"at  M\"unchen,
  Butenandstr.\  5-13, 81377 M\"unchen, Germany\\}
\author{K.\ Chadova} 
\affiliation{%
  Department  Chemie,  Physikalische  Chemie,  Universit\"at  M\"unchen,
  Butenandstr.\ 5-13, 81377  M\"unchen, Germany\\}
\author{D.\ K\"odderitzsch} 
\affiliation{%
  Department  Chemie,  Physikalische  Chemie,  Universit\"at  M\"unchen,
  Butenandstr.\ 5-13, 81377  M\"unchen, Germany\\}
\author{W.\ Bensch} 
\affiliation{%
Inst.\ f\"ur Anorgan.\ Chemie,  Universit\"at Kiel,
Olshausenstr.\ 40,  24098 Kiel, Germany\\}
\author{H.\ Ebert} 
\affiliation{%
  Department  Chemie,  Physikalische  Chemie,  Universit\"at  M\"unchen,
  Butenandstr.\ 5-13, 81377  M\"unchen, Germany\\}

\date{\today}
             
\begin{abstract}

The electronic, magnetic and transport properties of Fe intercalated 2H-TaS$_2$
have been investigated by means of the Korringa-Kohn-Rostoker  (KKR) method.
The non-stoichiometry and disorder in the system has been accounted for using
 the Coherent Potential Approximation (CPA) alloy theory.
 A pronounced influence 
of disorder on the spin magnetic moment  
has been found for the ferro-magnetically
ordered material. The same applies for the spin-orbit induced orbital magnetic
moment and magneto-crystalline anisotropy energy. 
The temperature-dependence of
the resistivity of disordered 2H-Fe$_{0.28}$TaS$_2$
investigated on the basis of the Kubo-St\v{r}eda formalism
in combination with the alloy analogy model has been found in very satisfying
agreement with experimental data. 
This also holds for the temperature dependent anomalous Hall resistivity
$ \rho_{\rm xy}(T) $. The role of thermally induced lattice vibrations
and spin fluctuations for the transport properties is discussed in detail.

\end{abstract}

\pacs{Valid PACS appear here}
\maketitle

\section{Introduction}

Transition metal dichalcogenides  are very attractive
materials both from a fundamental point of view as well
as concerning  potential technological applications.
 They are formed by well separated
 trilayers of the type MX$_2$
  (M: transition metal, X: S, Se, Te), that
determine their quasi-2D structure  leading to various interesting
physical properties. These are for instance 
strongly anisotropic transport properties \cite{FY87}
or
charge-density-wave (CDW) instabilities leading to the
 formation of CDW 
phases which can coexist at low temperature with superconductivity
\cite{WSM75,WW86,CNet01}. Many discussions in the literature 
concern the
opto-electronic properties of the semiconducting dichalcogenides 
and their relation 
to the features of the energy gap as width and  direct or indirect
character \cite{WY69,WKK+12}.

Most of the transition metal dichalcogenides 
are non-magnetic. However, they allow
the  intercalation of magnetic atoms or molecules 
between the X-M-X trilayers.
This leads to an interesting  
 class of magnetic materials
having quasi-2D properties \cite{PF80,PF80a,EMD+81,NIH+94} which can
be varied  via
 the amount and type of intercalated  atoms.
This is demonstrated in the 
present theoretical  work on  Fe intercalated 2H-TaS$_2$,
where the Fe atoms occupy the octahedral holes
between prismatic MX$_2$ trilayers (see Fig.\ \ref{fig:str}).
In fact,
 Fe intercalated 2H-TaS$_2$ 
can be seen as a prototype material
that 
 has been investigated
experimentally already some years ago
as the closely  related  intercalated 2H-structured systems
NbSe$_2$, TaSe$_2$, NbS$_2$, and TaS$_2$
with 
the magnetic
 $3d$-metals Mn, Fe, Co and Ni
as an  intercalate
 \cite{FBY77,vdBC68,ABCH70,LRI71}.  
As was shown, a
common feature of these materials is the  trend to create
 two-dimensional ordered $2 \times 2$ and 
           $\sqrt{3} \times \sqrt{3}$   structures
 within the $3d$-layer
 for  the
concentration  $x= 1/4$ and   $1/3$, respectively, of the  intercalate.

Depending on concentration and intercalation atom type,
these materials  exhibit for the ordered phase  
ferromagnetic (FM) \cite{PF80a} or 
anti-ferromagnetic (AFM) \cite{FBY77} magnetic order at low 
 temperatures. 
The AFM order  has been observed for the 
intercalated NbS$_2$ systems, e.g., 
Fe$_{1/3}$NbS$_2$,  
Co$_{1/3}$NbS$_2$,
and
Ni$_{1/3}$NbS$_2$, while in the case of 
TaS$_2$ based alloys intercalated
by $3d$-metals  FM order has been found \cite{PF80a}
for 
Cr$_{1/3}$TaS$_2$ ($T_{\rm C}=116$~K), 
Mn$_{1/3}$TaS$_2$ ($T_{\rm C}=70$~K),
and
Fe$_{1/3}$TaS$_2$ ($T_{\rm C}=35$~K). 
In the case of non-stoichiometric, disordered systems 
the Curie temperature depends in addition  on the
concentration of the $3d$ element \cite{EMD+81,NIH+94}. 
A strong dependence of $T_{\rm C}$ on the Fe concentration was found
in particular  for
Fe$_{x}$TaS$_2$ \cite{EMD+81}
with $T_{\rm C}= 90$,  $163$,     $90$,    and $55$~K
 for $x = 0.2$,  $0.26$,    $0.28$,      $0.34$, respectively.
At $x = 0.45$ 
the system becomes  AFM with a N\'eel temperature  $T_{\rm N} = 85$~K \cite{NIH+94}.

Ordered
Fe$_{1/4}$TaS$_2$ possesses 
a very pronounced out-of-plane magneto-crystalline anisotropy (MCA),
i.e.\ the easy axis is along the c-axis.
The extremely high   anisotropy
field $B_{\rm A}$ of about  $ 60$~T \cite{CLM+08,CKA+09},
that is by an  order of magnitude
higher than  observed for Mn$_{1/4}$TaS$_2$ and Mn$_{1/4}$NbS$_2$ \cite{PF80}
(about $0.5$~T),  
 leads  to an Ising-type
behavior of the Fe magnetic moment.

The transport properties
of magnetic  intercalated transition metal dichalcogenides
 have been investigated by various authors.
Examples for this 
are the electrical resistivity and anomalous Hall effect (AHE) \cite{FBY77}
studied for alloys based on 2H-NbS$_2$.
The magneto-transport and superconducting properties of 
disordered
Fe intercalation compounds
based on 
NbSe$_2$, TaSe$_2$ and TaS$_2$ 
have been investigated  by Whitney {\em et al.} \cite{WFC77}
for the dilute regime ($x < 0.1$).
Recently, the results of 
magneto-transport measurements for Fe$_{1/4}$TaS$_2$ have been
reported by Morosan {\em et al.} \cite{MZL+07}. This material shows a
strong anisotropy concerning the magnetization as well as the resistivity. 
Checkelsky {\em et al.} \cite{CLM+08} focused on the temperature dependence of the 
resistivity of Fe$_{1/4}$TaS$_2$. These authors  
showed that the characteristics  of the AHE for 
$T>50$~K cannot be 
explained by the Karplus-Luttinger or Berry-phase mechanism.
Instead it was concluded that it is governed
by  scattering processes
connected with temperature induced phonons and magnons. 
These authors also discussed the unconventional
behavior of the magneto-resistance as a function of temperature.

A very large magneto-resistance (MR) has been found 
recently  for disordered
Fe$_{0.28}$TaS$_2$ single crystal by  Hardy {\em et al.} \cite{HCM+15}, which is
nearly $100$ times stronger than that 
observed for the ordered Fe$_{1/4}$TaS$_2$ 
compound. The authors point out the crucial role 
of
spin disorder and strong spin-orbit coupling  in the system,
that can be exploited  to create materials with large MR. 

In the present work we will focus on TaS$_2$
intercalated with Fe
as a prototype material for 
magnetic  intercalated transition metal dichalcogenides.
While several experimental studies on the magnetic and transport
properties of this system can be found in the literature,
only few theoretical investigations \cite{DZvB+89,KKK+11}
    have been done so far
with the focus on its   electronic structure
assuming an ordered state.

\section{Computational details}

\label{SEC:Computational-scheme}

The present theoretical
investigations on the  magnetic and transport properties of Fe
intercalated 2H-TaS$_2$ are based on  first-principles electronic
structure calculations which have been performed using the fully
relativistic KKR Green function method.\cite{Ebe00,EKM11}
The 
combination with the coherent potential approximation (CPA) 
alloy theory
allowed to deal with disorder on the Fe sub-lattice
as well as the impact of finite temperatures 
on transport properties on the basis of   the alloy analogy model (see below).
The self-consistent calculations have been done in the
framework of the local approximation to spin density
functional theory (LSDA) using the parametrization for
the exchange and correlation potential as given by
 Vosko {\em et al.} \cite{VWN80}.
Correlation effects going beyond the level of LSDA
have been  accounted for by means 
of the  LDA+U scheme \cite{SLP99,YAF03,EPM03}
using for Fe the parameters 
 $U = 4.5 $~eV and $J = 0.7$~eV \cite{KKK+11} throughout.

\begin{figure}[h]
\includegraphics[width=0.3\textwidth,angle=0,clip]{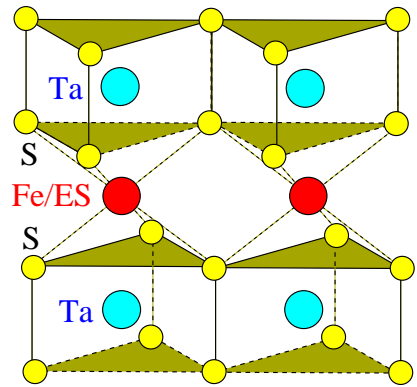}\\
\caption{\label{fig:str}
 The structure of  the investigated 
 Fe intercalated 2H-TaS$_2$ system showing the position 
occupied by
 the Fe atoms and empty spheres (ES) in between the S-Ta-S trilayers
according to the Fe concentration.
 }  
\end{figure}

The  parameters  specifying the structure 
of   Fe intercalated 2H-TaS$_2$
as shown in Fig.\ \ref{fig:str}          
have been taken from the experiment
\cite{NIH+94,MZL+07}.
%
%
For the angular momentum
expansion of the Green function a cutoff of
$l_{max} = 3$ was applied.  
This is in particular important for the transport calculations
that were based on the  Kubo-St\v{r}eda  formalism \cite{SS77}
that among others  gives access to the 
AHE.\cite{LKE10b} To deal with the temperature dependence
of the transport  properties a scheme has been used
that is based on the alloy analogy model and 
 that accounts for thermal 
lattice vibrations \cite{KCME13} as well as spin fluctuations.\cite{EMC+15}
Within this approach the spin fluctuations are represented by
a temperature-dependent quasi-static spin configuration
corresponding to the adiabatic approximation.\cite{AKH+96}
The spin configuration used as an input for the transport calculations 
may be determined for example by performing Monte Carlo
simulations or deduced from the experimental 
temperature-dependent magnetization.\cite{EMC+15}
Here, the later scheme has been applied using
experimental data from Ref.\ \onlinecite{HCM+15}.

\section{Results}
\subsection{Electronic structure}

At low temperatures, pure 2H-TaS$_2$  exhibits 
a charge density wave (CDW) instability \cite{GSR+11} driven by the
Fermi surface nesting mechanism,  bringing the system into the CDW state
below $T_{\rm CDW} = 80$~K  \cite{WW86}. 
Intercalation with  Fe obviously strongly
modifies the Fermi surface due to Fe-related  electronic states resulting in a
shift of the Fermi energy and a smearing of the electronic bands 
due to the  disorder within the Fe layers. This suppresses  the CDW instability
and the intercalated system shows conventional  metallic behavior. 

\begin{figure}[h]
\includegraphics[width=0.43\textwidth,angle=0,clip]{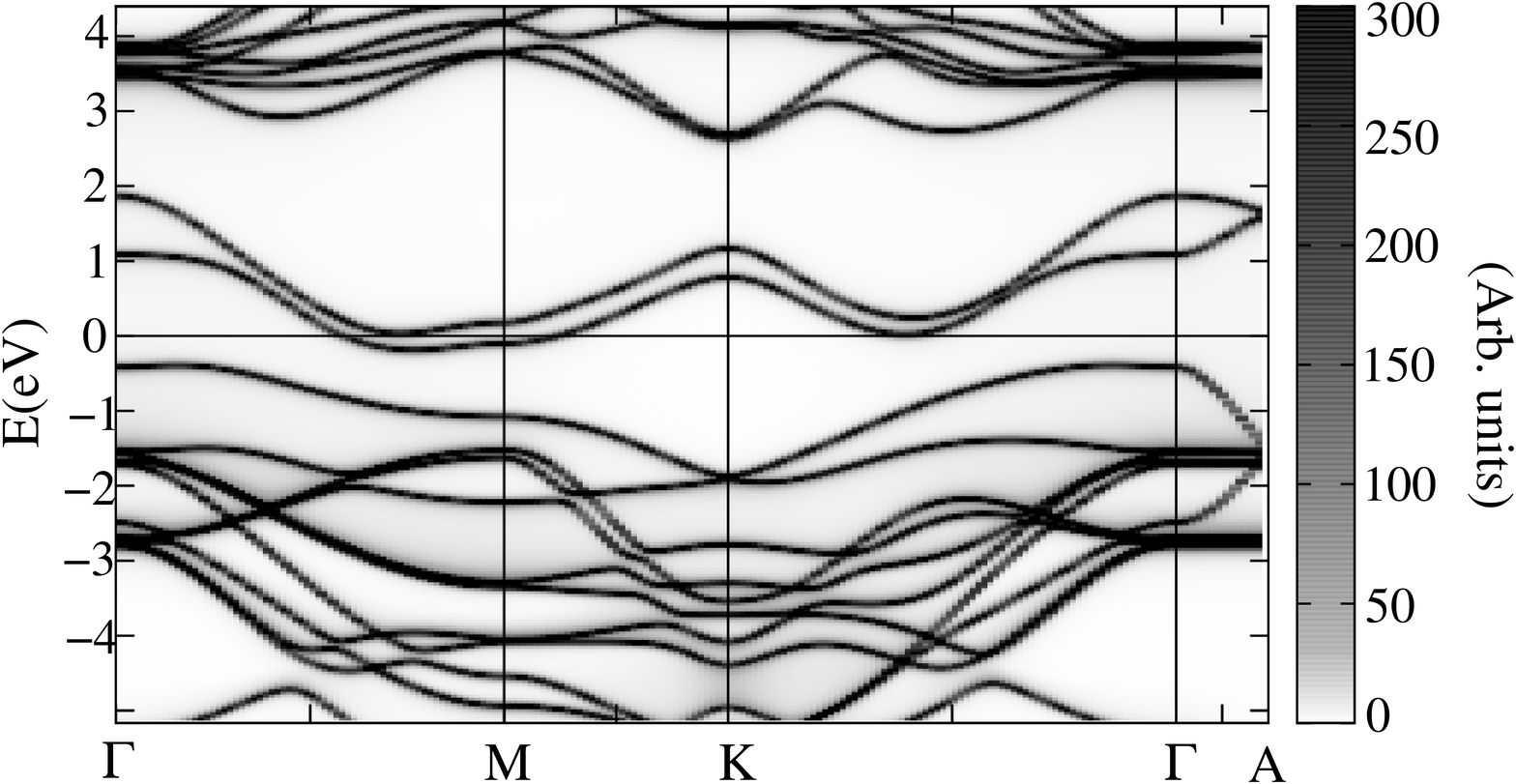}\;(a) \\
\includegraphics[width=0.43\textwidth,angle=0,clip]{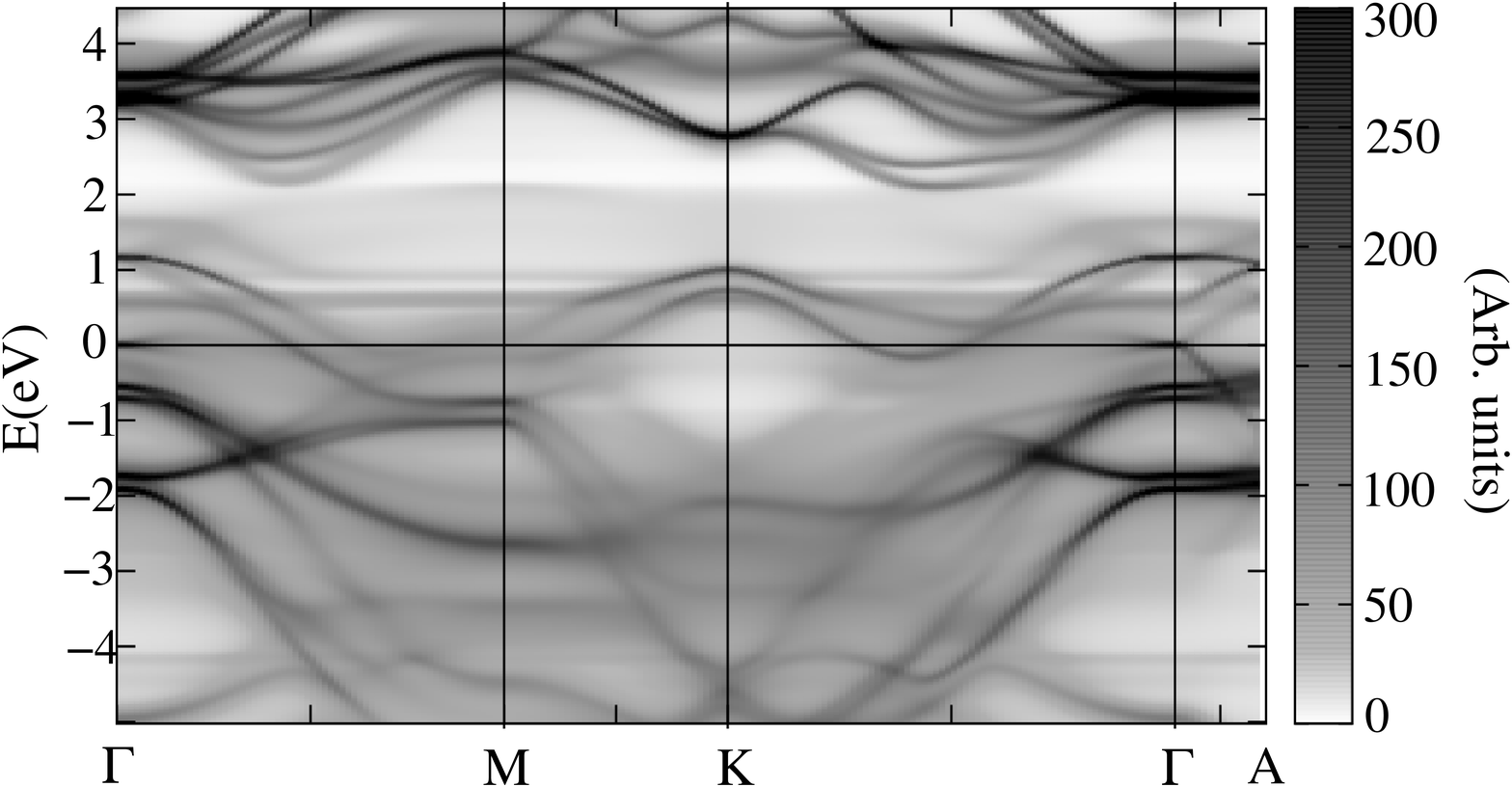}\;(b) \\
\includegraphics[width=0.43\textwidth,angle=0,clip]{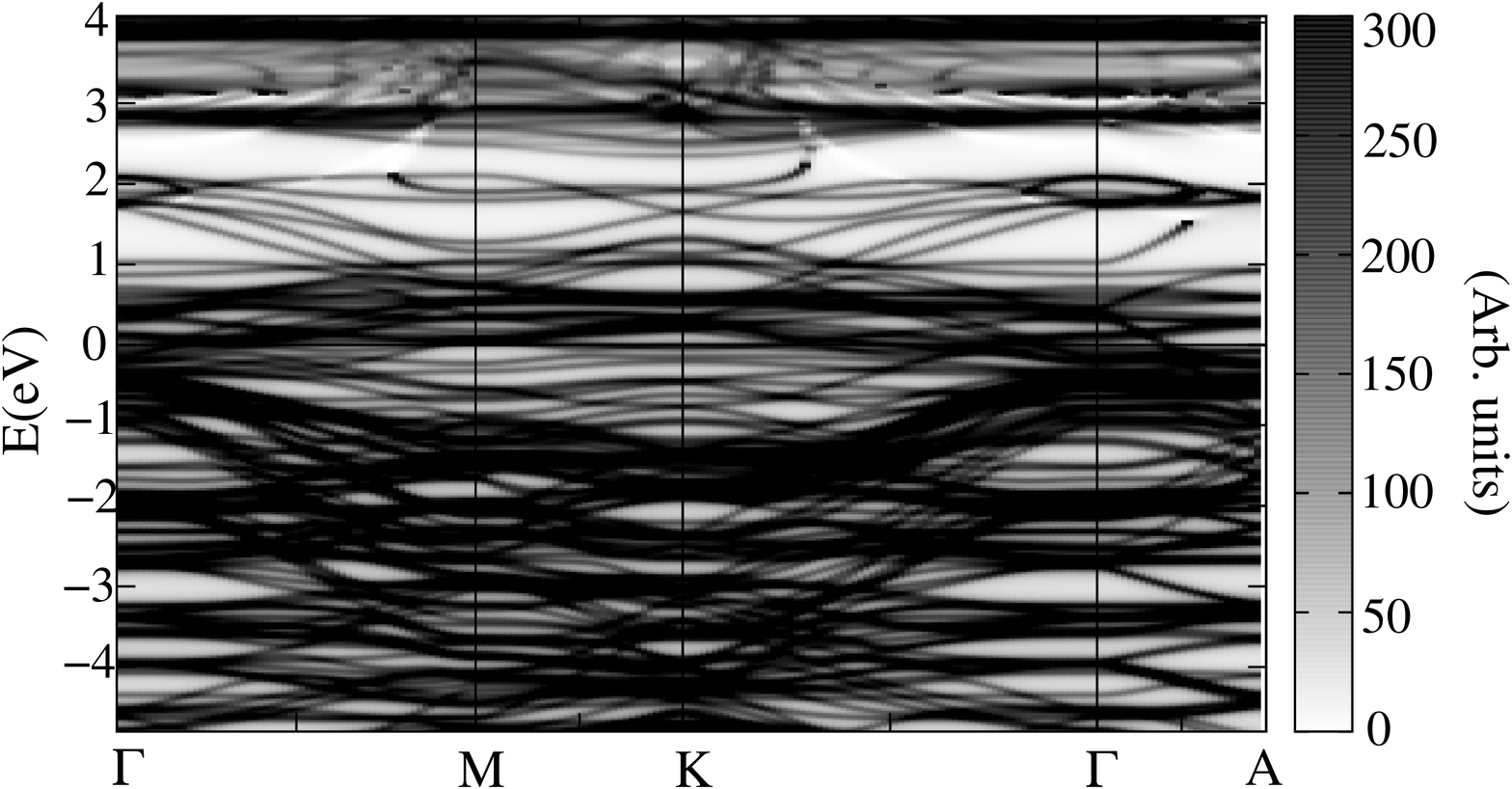}\;(c)
 \caption{\label{fig:BSF2H}
Spin-integrated
 Bloch spectral function $A(\vec{k},E)$
   along high symmetry directions of the  Brillouin zone calculated for: 
(a)
   2H-TaS$_2$, within LSDA; 
(b) ferromagnetic disordered Fe$_{0.25}$TaS$_2$, 
   within LSDA+U; and 
(c) ferromagnetic ordered Fe$_{1/4}$TaS$_2$, within
   LSDA+U. In the case of ordered compounds, TaS$_2$ and ordered
   Fe$_{1/4}$TaS$_2$, the BSF was calculated for a small imaginary part
   $\Im(E)  = 0.001$~Ry of the energy $E$.}   
 \end{figure}

These findings are reflected by the results of calculations on the 
electronic structure.
Fig.\ \ref{fig:BSF2H} shows 
the corresponding spin-integrated
 Bloch spectral function (BSF) $A(\vec{k},E)$
 calculated for 
   2H-TaS$_2$, 
 disordered Fe$_{0.25}$TaS$_2$,
as well as  ordered Fe$_{1/4}$TaS$_2$.
%
%
To avoid the use of an extremely fine mesh of $\vec k$- and $E$-points
these calculations were done 
using a  small imaginary part
 of the energy $E$
in the case of the ordered compounds  TaS$_2$ and 
   Fe$_{1/4}$TaS$_2$.
The resulting BSF for TaS$_2$ 
corresponds essentially to the dispersion relation
$E(\vec k)$ given by 
Blaha \cite{Bla91} with the differences 
primarily to be ascribed to the impact of the 
spin-orbit coupling that is accounted for within the present 
fully relativistic calculations.
Intercalation of Fe leads for the disordered case, 
apart from the exchange splitting,
primarily to a broadening of the bands as can be seen from 
Fig.\ \ref{fig:BSF2H} b).
For ordered
   Fe$_{1/4}$TaS$_2$, on the other hand,
   the band structure gets much more complex due to the 
   increase of the size of the unit by a factor 
   of 4 and the occurrence of new Fe-related states
(see Fig.\ \ref{fig:BSF2H} c)).

\begin{figure}[h]
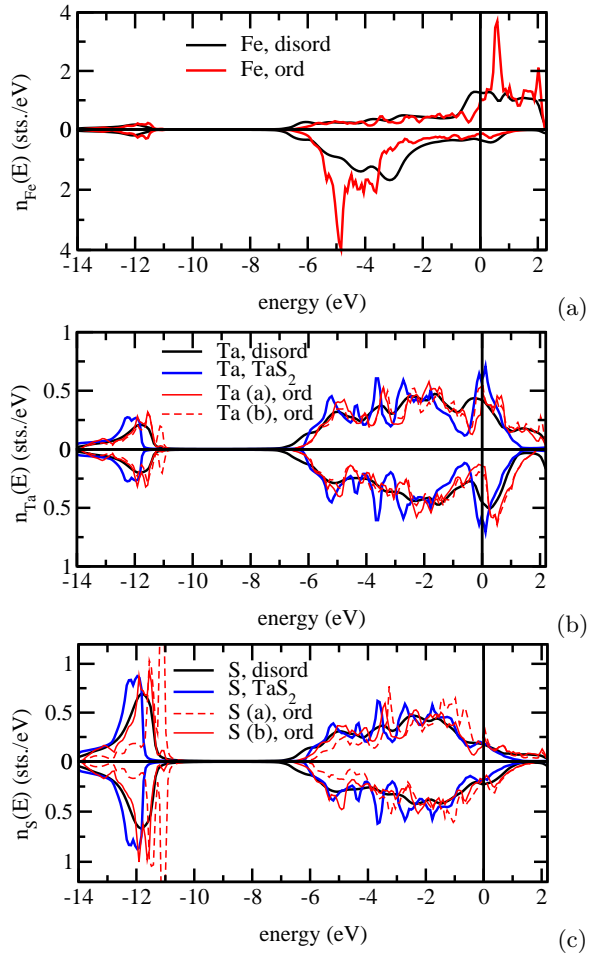

\includegraphics[width=0.4\textwidth,angle=0,clip]{CMP_DOS_FeTaS2_vs_TaS2_Fe_mod.eps}\;(a)
\includegraphics[width=0.4\textwidth,angle=0,clip]{CMP_DOS_FeTaS2_vs_TaS2_Ta_mod.eps}\;(b)
\includegraphics[width=0.4\textwidth,angle=0,clip]{CMP_DOS_FeTaS2_vs_TaS2_S_mod.eps}\;(c)
\caption{\label{fig:DOS2H} Fe (a), Ta (b) and S (c) DOS in pure TaS$_2$
  (within LSDA) and  ferromagnetic
disordered  Fe$_{0.25}$TaS$_2$  as well as 
  ordered Fe$_{1/4}$TaS$_2$
   (within LDA+U).
    In the cases of Ta (b) and S (c)
  the thin solid and dashed red lines represent the DOS for two
  in-equivalent sites, a and b, in ordered Fe$_{1/4}$TaS$_2$.  }  
\end{figure}

The various features in the BSF in  Fig.\ \ref{fig:BSF2H}
are also reflected by the component and
spin-resolved density of states (DOS) shown 
in  Fig.\ \ref{fig:DOS2H}
for  Fe (a), Ta (b) and S (c)  in pure TaS$_2$,
 disordered  Fe$_{0.25}$TaS$_2$ as well as 
  ordered Fe$_{1/4}$TaS$_2$. 
%
%
The  rather broad Fe energy bands 
(Fig.\ \ref{fig:DOS2H} (a))
 indicate a  significant hybridization with
the electronic states of the host atoms, especially with the 
neighboring  S atoms. This can also
be concluded from the partial   DOS of S, 
shown in Fig.\ \ref{fig:DOS2H} (c). 
Obviously, disorder 
in the Fe sub-lattice leads for Fe itself to quite pronounced 
changes when compared to the ordered case; apart from 
a broadening and resulting smearing of the spectral features
in the Fe DOS one notices  a substantial redistribution of
spectral weight together with an apparent change
in the exchange splitting. For the S and Ta sublattices, on the other side,
disorder leads first of all to a smearing of the DOS curve.
Interestingly the impact of disorder for Ta is at least as pronounced
as for the S, being next-nearest neighbor to Fe 
(see Fig.\ \ref{fig:str}).

\subsection{Magnetic moments}

\begin{figure}[h]
\includegraphics[width=0.4\textwidth,angle=0,clip]{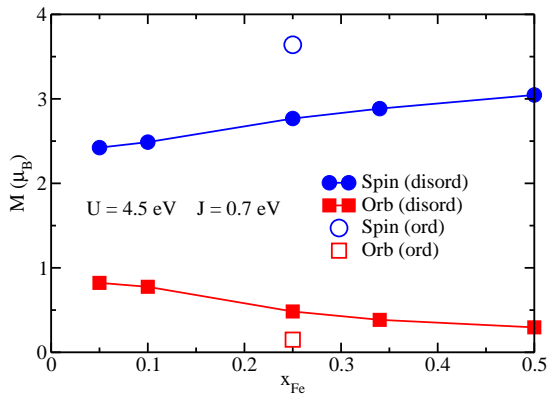}
\caption{\label{fig:MAG-MOM} Spin (circles) and orbital (squares) 
magnetic moments of Fe
 in ferromagnetic 2H-Fe$_{x}$TaS$_2$ as a function of
the Fe concentration. Results for the disordered state are given 
by full symbols while those for ordered  2H-Fe$_{1/4}$TaS$_2$  
are given by open symbols.}  
\end{figure}

Calculations of the magnetic moments 
for ferromagnetic disordered 2H-Fe$_{x}$TaS$_2$
have been done first
using the GGA approach for the treatment of exchange
and correlation. As this led to results that were too low when
compared to experiment (see below) the LDA+U method
was used instead with the corresponding parameters given in section
\ref{SEC:Computational-scheme}.   
As can be seen from 
Fig.\ \ref{fig:MAG-MOM} 
 this led for disordered
  2H-Fe$_{x}$TaS$_2$ to a spin magnetic moment 
that increases nearly monotonously
from about 2.42 to 3.05~$ \mu_{\rm B}$ when the Fe concentration $x$ 
increases from 0.05 to 0.5. 
%
%
On the other hand, the 
relatively large
 spin-orbit induced orbital moment decreases from about 
0.82 to 0.29~$ \mu_{\rm B}$.
In the case of ordered 2H-Fe$_{1/4}$TaS$_2$  quite pronounced changes
 compared to the
disordered state are found; upon ordering the 
spin moment increases from    2.76 to 3.64~$ \mu_{\rm B}$ while the 
orbital moment decreases from 0.48 to 0.16~$ \mu_{\rm B}$.
For the GGA calculations the corresponding changes are
 1.93  to  3.1~$ \mu_{\rm B}$ for the spin moment and 
 0.12  to  0.11~$ \mu_{\rm B}$ for the orbital moment, respectively.
%
%
The increase of the spin magnetic with ordering is quite common and can
often be associated with the DOS of the ordered state that is
more structured and less broad than for the disordered state
(see Fig.\ \ref{fig:DOS2H}).
For the same reason one has usually also an increase of the orbital magnetic
moment. On the other hand,  the orbital magnetic moment depends much more on
the details of the
electronic structure in the vicinity of the Fermi energy than the spin magnetic moment.
This together with the relatively small width of the Fe sub-band 
seems to be the reason for the observed pronounced decrease for the orbital moment
upon ordering.

The LDA+U-based results differ quite substantially from that of 
Ko {\em et al.} \cite{KKK+11} who used the GGA+U scheme
with the values for  $U$ and $J$ as given in section
\ref{SEC:Computational-scheme}  
and  spin-orbit coupling treated as a perturbation.
This approach led  for the spin
and orbital magnetic moment to the values 2.95 and 1.0~$ \mu_{\rm B}$,
 respectively.
These results imply that in particular the orbital magnetic moment depend
quite sensitively on the treatment of exchange and correlation as well as
spin orbit coupling.

\medskip

Experimental work  on ordered 2H-Fe$_{1/4}$TaS$_2$
led to a total magnetic moment per Fe atom of 
3.9~$ \mu_{\rm B}$ and a Curie temperature of 
$T_{\rm C}=160$~K.\cite{MZL+07}
This is in close agreement  with more recent work 
by  Checkelsky {\em et al.} \cite{CLM+08} ($T_{\rm C}=160$~K)
as well as  Choi {\em et al.}\cite{CKA+09}.
The later authors  did measurements
on ordered samples obtained as grown (AG) and by
slow cooling (SC) 
leading to a total moment of   4.0~$ \mu_{\rm B}$
and a Curie temperature of 
$T_{\rm C}=156$ and $159$~K, respectively.
Additional measurements on quenched (Q) disordered
 samples led to a substantially
lower Curie temperature of  $T_{\rm C}=104$~K.

On the basis of  XMCD measurements at the  L$_{2,3}$-edges of Fe
     the ratio of the 
spin ($m_{\rm spin}$)  to  the orbital ($m_{\rm orb}$)  moment
 has been estimated 
to be  $m_{\rm orb}/m_{\rm spin} = 0.33$ \cite{KKK+11}.
Assuming an ionic configuration for Fe 
with a spin moment of  4.0~$ \mu_{\rm B}$ an
orbital moment of   1.33~$ \mu_{\rm B}$ 
was suggested.
Using instead  the value 4.0~$ \mu_{\rm B}$
  for the total moment as given above an 
orbital moment of   1.0~$ \mu_{\rm B}$ results from 
the analysis of the XMCD measurements.

\medskip

When comparing the experimental results with the
theoretical ones given in Fig.\ \ref{fig:MAG-MOM} 
one finds that  the spin magnetic moment calculated for the  
2H-Fe$_{1/4}$TaS$_2$ compound is in a good agreement with
the experimental data  while the  orbital one
seems to be too low.
This might be to a large extent  due to the pronounced
dependence of the   orbital magnetic moment
on the computational details that was discussed 
 above.

\subsection{Magneto-crystalline anisotropy }

\begin{figure}[h]
\includegraphics[width=0.4\textwidth,angle=0,clip]{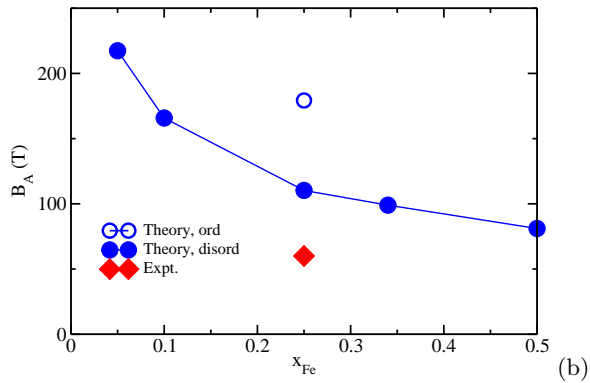}\;(b)
\caption{\label{fig:MCA_LDAU} 
Magneto-crystalline anisotropy  in the Fe-intercalated TaS$_2$ vs Fe
concentration. The  diamond represents the MCA energy measured 
at 2~K 
\cite{CLM+08,CKA+09} (anisotropy field  $B_{\rm A} = 60$~T) }  
\end{figure}

The magneto-crystalline anisotropy  energy $\Delta E_{\rm MCA} $
defined as the difference in energy for the in-plane and out-of-plane
orientation of the magnetization has been calculated by means of
the so-called torque method \cite{SSB+06}. 
$\Delta E_{\rm MCA}(x) $ determined this way 
for disordered
Fe$_{x}$TaS$_2$ 
 as a function of the Fe concentration $x$
and expressed  with respect to the formula unit (see below) 
increases monotonously with $x$.
To compare with experimental data $\Delta E_{\rm MCA}(x) $
was converted into an anisotropy field $B_{\rm A} =\Delta E_{\rm MCA}  / M_{\rm f.u.} $,
where the magnetic moment per formula unit ($M_{\rm f.u.} $)  
is given by that per Fe atom via 
$M_{\rm f.u.} = x M_{\rm Fe}$.
Using this representation
for the magneto-crystalline anisotropy $B_{\rm A}(x) $  decreases monotonously with $x$ as can be seen
in Fig.\ \ref{fig:MCA_LDAU}. 
%
%
For $x=0.25$ calculations have been done in addition 
assuming an ordered compound  Fe$_{1/4}$TaS$_2$.
As often found, ordering increases the anisotropy field in a quite
pronounced way. Astonishingly, this consequence of ordering 
for $  \Delta E_{\rm MCA}(x) $ is  reversed when compared to 
the situation for 
the orbital magnetic moment. 
Comparison of the theoretical result for 
$B_{\rm A} $ with the experimental value for $x=1/4$
confirms the large anisotropy field found in experiment.
Nevertheless, the theoretical value 
for the ordered  as well as for the disordered state are well above 
the experimental value. This indicates again that a certain amount
of disorder may be present in the sample investigated.

\subsection{Electrical resistivity}

\begin{figure}[th]
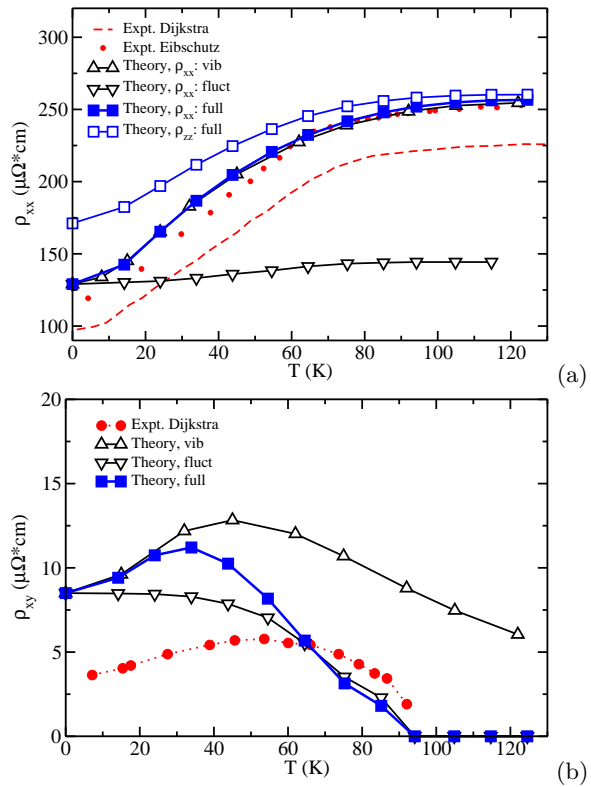

\includegraphics[width=0.4\textwidth,angle=0,clip]{rho_xx_Modified.eps}\;(a)\\
\includegraphics[width=0.4\textwidth,angle=0,clip]{rho_xy_Modified.eps}\;(b)
\caption{\label{fig:rhoxx}
(a): 
Temperature-dependent longitudinal resistivities,
 $\rho_{\rm xx}(T)=\rho_{\rm yy}(T)$ and  $\rho_{zz}(T)$ (filled and open squares)
 for ferromagnetic  disordered
  2H-Fe$_{0.28}$TaS$_2$.
  The dots \cite{EMD+81} and the dashed
  line \cite{DZvB+89} represent corresponding experimental data. 
(b)
Temperature-dependent transverse resistivity, $\rho_{\rm xy}(T)$.
Results for  $\rho_{\rm xx}(T)$ and $\rho_{\rm xy}(T)$ that were obtained  accounting
 only for  lattice vibrations and  spin fluctuations
 are  represented  in both cases by up and down triangles, respectively.
  }  
\end{figure}

The temperature-dependent 
 longitudinal electrical resistivities,
 $\rho_{\rm xx}(T)=\rho_{\rm yy}(T)$ and $\rho_{zz}(T)$, 
 calculated for ferromagnetic
disordered  2H-Fe$_{0.28}$TaS$_2$ using the scheme described
in section \ref{SEC:Computational-scheme}
are shown in
Fig.\ \ref{fig:rhoxx} (a). 
%
%
For the  in-plane longitudinal electrical resistivity $\rho_{\rm xx}(T)$ 
results are given in addition  that account only for the lattice vibrations (vib)
and  spin fluctuations (fluct), respectively.
For $T=0$~K all individual $\rho_{\rm xx}^{\rm }(T)$ curves represent the
residual resistivity due to the random disorder in the Fe sub-lattice. 
Keeping the magnetic spin configuration fixed
and dealing only with the impact of
  thermal  lattice vibrations on the resistivity
a pronounced and monotonous increase with temperature is observed
for the 
 corresponding
 curve   $\rho_{\rm xx}^{\rm vib}(T)$ (up triangles). 
Considering  on the other hand
$\rho_{\rm xx}^{\rm fluct}(T)$ (down triangles) 
 that accounts for the  spin fluctuations only 
 one finds a rather weak increase
 with temperature.
Accounting for both temperature dependent scattering mechanisms 
simultaneously one finds for the resulting curve    $\rho_{\rm xx}^{}(T)$
(full squares) only minor changes 
compared to    $\rho_{\rm xx}^{\rm vib}(T)$. 
As found before for ferromagnetic bcc-Fe \cite{EMC+15},
this finding clearly shows 
that  $\rho_{\rm xx}^{\rm fluct}(T)$ and  $\rho_{\rm xx}^{\rm vib}(T)$ are not additive,
i.e.\ the Matthiesen rule is violated.
Comparing the calculated  $\rho_{\rm xx}^{}(T)$ curve 
with the available experimental data  
\cite{EMD+81,DZvB+89} one can see that the 
experimental and theoretical  residual resistivities
agree rather well. The fact that the theoretical values are 
somewhat higher may indicate that the assumption of fully random disorder
on the Fe sub-lattice for the calculations 
is not fully justified; i.e.\ there might
be some order present in the 
experimentally  investigated samples.

The various curves in 
Fig.\ \ref{fig:rhoxx} (a) clearly show that the 
thermal lattice vibrations are primarily responsible 
for the temperature dependence of  $\rho_{\rm xx}^{}(T)$ 
including the change of its slope close to the Curie temperature.
For that reason this peculiar behavior cannot
be ascribed to the temperature induced magnetic disorder
as it was assumed before.\cite{EMD+81,DZvB+89}
Obviously, the temperature dependence of  $\rho_{\rm xx}^{}(T)$
has primarily to be ascribed to the 
increasing  smearing of the energy bands 
with temperature
due to electron scattering caused by lattice vibrations. 
For low temperature, this results in a rather fast increase of the
resistivity due to a corresponding
increasing cross section for the inter-band scattering. 
For higher temperatures, a 
saturation of the number of channels for the inter-band scattering 
seems to occur leading finally to a rather weak 
increase of  $\rho_{\rm xx}(T)$ with temperature.

As can be seen in Fig.\ \ref{fig:rhoxx} (a),
 the calculated out-of-plane resistivity 
 $\rho_{zz}(T)$  for $T = 0$~K 
 is larger than its in-plane counterpart  $\rho_{\rm xx}(T)$
 reflecting the 2D-character of the system.
The finding that this difference is relatively weak is due to the
disorder in the Fe sub-lattice that causes in both cases a large
residual resistivity.
 With increasing temperature this 
difference diminishes monotonously being nearly absent for
the highest temperature considered here  ($T = 125$~K).
This behavior of the resistivity
has been observed and discussed 
before for example for the compounds
 Nb$_3$Sn and Nb$_3$Sb.\cite{CCH67,FW76,AC81}  
The fact that the  $\rho_{zz}(T)$-curve
is not  shifted rigidly against that for 
 $\rho_{\rm xx}(T)$  clearly shows once more that the
 contributions to the resistivity
due to chemical and thermal disorder are not
simply additive.
 This  non-additive behavior 
together with the fact that 
the temperature dependence 
for  $\rho_{xx}(T)$ as well as  for  $\rho_{\rm zz}(T)$ 
is obviously dominated by the thermal lattice vibrations
leads obviously to the decreasing anisotropy of the 
resistivity with increasing  temperature 
seen in  Fig.\ \ref{fig:rhoxx} (a).

\medskip

The temperature-dependent 
 transverse  electrical resistivity $\rho_{\rm xy}(T)$ 
for
disordered  2H-Fe$_{x}$TaS$_2$
 is
 shown in  Fig.\ \ref{fig:rhoxx} (b). 
For $T=0$~K the various theoretical  curves are determined only 
by the chemical disorder on the Fe sub-lattice and for that 
reason all coincide. The value
 $\rho_{\rm xy}(0)$ may be decomposed into its intrinsic or coherent 
  and extrinsic or incoherent contribution \cite{LKE10b,TKD12}.
  The later part represents in particular all so-called 
  skew scattering contributions.
Keeping the spin configuration fixed to that for $T=0$~K
(corresponding to a collinear ferromagnetic ordering)
and accounting for the lattice vibrations only the resulting
$\rho_{\rm xy}^{\rm vib}(T) $  
rises with temperature $T$ and shows a broad maximum around $40$~K.
For higher temperatures 
$\rho_{\rm xy}^{\rm vib}(T) $  decreases with $T$ as was found before 
for example in the case of pure Ni in the regime below
the Curie temperature.\cite{KCME13}
On the other hand, if the lattice structure is kept undistorted 
and spin fluctuations are accounted for 
$\rho_{\rm xy}^{\rm fluct}(T) $   monotonously decreases with $T$
until the Curie temperature is reached and   $\rho_{\rm xy}^{\rm fluct}(T) $  
vanishes as the thermally averaged z-component
of the magnetization $\langle M_z \rangle_T$ vanishes.
If both scattering mechanism are accounted for one finds again that these
do not act in a simple additive way.
Nevertheless, from the behavior of the individual curves
$\rho_{\rm xy}^{\rm vib}(T) $ and 
$\rho_{\rm xy}^{\rm fluct}(T) $  one may expect the 
behavior of the transverse resistivity 
$\rho_{\rm xy}^{}(T) $  that account for the lattice vibrations and 
spin fluctuations simultaneously.
When comparing these results 
with the corresponding experimental data, 
a reasonable agreement is found. In particular 
concerning the presence of a maximum for 
$\rho_{\rm xy}^{}(T) $  with temperature.
The individual curves 
$\rho_{\rm xy}^{\rm vib}(T) $ and 
$\rho_{\rm xy}^{\rm fluct}(T) $ 
clearly demonstrate the importance of extrinsic contributions 
to $\rho_{\rm xy}^{}(T) $  supporting the conclusions 
of 
Checkelsky {\em et al.} \cite{CLM+08}  concerning their role 
for 
 Fe$_{x}$TaS$_2$ mentioned in the introduction.

\section{Summary}

Using the KKR-CPA band structure method for disordered systems
 the electronic structure as well as magnetic
 and transport properties  of ferromagnetic 
ordered and disordered  2H-Fe$_{ x}$TaS$_2$
have been investigated.
By means of the fully relativistic of the KKR-CPA 
the spin-orbit induced 
orbital magnetic
moment and magneto-crystalline anisotropy energy
could be calculated in particular.
The various magnetic properties were found in reasonable agreement
with available experimental data with clear indications
for  the strong impact of the degree of order on the
Fe sub-lattice of the system.
In addition,  a  prominent role of correlation effects
was found that were ascribed to relatively narrow with
of the Fe-related bands. 
The temperature-dependence of the longitudinal  resistivity
as well as transverse anomalous Hall resistivity
was studied using the alloy analogy model 
within the framework of the  Kubo-St\v{r}eda formalism.
The results obtained for disordered 2H-Fe$_{0.28}$TaS$_2$
were  found in very satisfying
agreement with experimental data. 
Additional calculations accounting for  thermally induced
lattice vibrations and spin fluctuations 
individually
clearly showed that their contribution to the resistivity
is not additive and that the 
 temperature-dependence of the longitudinal  resistivity
is nearly exclusively determined by the lattice vibrations.

\acknowledgements

 Financial support by the    Deutsche
  Forschungsgemeinschaft (DFG) via the priority programs SPP 1415
and SPP 1538
is thankfully acknowledged.


%

\end{document}